\documentclass{article}

\usepackage{cite}
\usepackage{amsmath}
\usepackage{amsfonts}
\usepackage{amssymb}
\usepackage{bm}
\usepackage{bbm}

\begin{document}

\vspace*{1cm}

\begin{center}
\hrule 
\vspace*{0.08cm}
\hrule 
{\large \bf \rule[-3mm]{0mm}{8mm}
Two--Photon Decays Reexamined: 
Cascade Contributions and Gauge Invariance}\\[2ex]
\hrule 
\vspace*{0.08cm}
\hrule
\vspace*{0.2cm}
\normalsize
Ulrich D. Jentschura\\[1ex]
\small
{\it Max--Planck--Institut f\"ur Kernphysik,
Postfach 10 39 80, 69029 Heidelberg, Germany}\\
{\it and Institut f\"ur Theoretische Physik,
Philosophenweg 16, 69120 Heidelberg, Germany}\\[4ex]
\normalsize
\begin{minipage}{12.5cm}
The purpose of this paper is to calculate the 
two-photon decay rate corresponding to the 
two-photon transitions $nS \to 1S$ and
$nD \to 1S$ in hydrogenlike ions with a low 
nuclear charge number $Z$ (for principal 
quantum numbers $n = 2, \dots, 8$).
Numerical results are obtained within a nonrelativistic framework,
and the results are found to scale approximately as
$(Z\alpha)^6/n^3$, where $\alpha$ is the fine-structure constant.
We also attempt to clarify a number of subtle 
issues regarding the treatment of the 
coherent, quasi-simultaneous emission of the two photons 
as opposed to one-photon cascades.
In particular, the gauge invariance of the decay rate 
is shown explicitly.\\[1ex]

\end{minipage}
\end{center}

\section{Introduction} 

The subject of the current paper is the two-photon decay rate
of excited atomic states, interpreted as the imaginary
part of the two-loop self-energy. We follow our previous 
investigation reported in Ref.~\cite{Je2007} and augment 
the analysis by treating the decay rate in both length and 
velocity gauges. Special emphasis is placed on the role of 
singularities, infinitesimally displaced from the integration
contours for the photon energy integrations,
which are generated by bound-state poles of lower energy
than the reference state (in the sense of the two-loop self-energy).
The reference state is equivalent to the initial state of the 
two-photon decay process. 
A good quantitative understanding of the 
two-photon decay processes from highly excited 
hydrogenic bound states is important for 
astrophysics, as emphasized in a recent paper by Chluba and 
Sunyaev~\cite{ChSu2007}. As the physics of the process is 
in principle well known and has been discussed in a
previous fast-track communication~\cite{Je2007}, we see no obstacle to going 
{\em in medias res} with the analysis.

Our purpose here, in addition to providing numerical 
data concerning the $D \to S$ transitions, is to clarify 
the role of cascades of one-photon decays through so-called 
resonant intermediate states, which are addressed using 
concepts developed in field theory~\cite{Ri1972,BaKoSt1998}.

Natural units with $\hbar = c = \epsilon_0 = 1$,
i.e.~$e^2 = 4 \pi \alpha$, are used throughout this paper,
which is organized as follows. In Sec.~\ref{gauge_invariance}, 
the gauge invariance of the two-photon decay rate, as derived 
from the two-loop self-energy, is reanalyzed. 
In Sec.~\ref{numerical_results}, 
numerical results for $nD \to 1S$ transitions are presented;
these were not treated in the previous paper~\cite{Je2007}.
A discussion of our results, including a comparison to 
previous investigations of two-photon decay from highly excited 
states (see Refs.~\cite{Fl1984,FlScMi1988,CrTaSaCh1986})
is given in Sec.~\ref{discussion}. Cascade contributions
are analyzed in Sec.~\ref{cascades}. Conclusions are drawn in 
Sec.~\ref{conclusions}.

%
%
\section{Gauge Invariance} 
\label{gauge_invariance}

We start by considering the two-photon self-energy 
for a reference state $| \phi_i \rangle$ in a hydrogenlike ion,
as derived from nonrelativistic quantum electrodynamics
(NRQED). In the velocity gauge, the interaction Hamiltonian 
of the quantized electromagnetic field with the 
electron is given by
\begin{equation}
H_I = - \frac{e}{2m}\, 
\left( \vec{p}\cdot \vec{A} +
\vec{A}\cdot \vec{p} \right) +
\frac{e^2 \vec{A}^2}{2 m}\,,
\end{equation}
where $\vec{A}$ is the vector potential of the quantized 
electromagnetic field.

The well-known expression (see, e.g., Refs.~\cite{Pa2001,Je2007})
for the two-loop self-energy reads ($\omega_1$ and $\omega_2$ denote
the energies of the two virtual quanta)
\begin{equation}
\label{NRQED_velocity}
\Delta E^{(2)}_i = \lim_{\epsilon \to 0}
\left( \frac{2 \alpha}{3 \pi m^2} \right)^2
\int_0^{\Lambda_1} d\omega_1 \, \omega_1 
\int_0^{\Lambda_2} d\omega_2 \, \omega_2 \, 
f_\epsilon(\omega_1, \omega_2) = 
{\rm Re} \Delta E^{(2)}_i 
- {\rm i} \, \frac{\delta \Gamma_i^{(1)}}{2}
- {\rm i} \, \frac{\Gamma_i^{(2)}}{2}\,.
\end{equation}
Here, ${\rm Re} \Delta E^{(2)}_i$ is the real part of the 
energy shift, which gives rise, in particular, to the 
so-called two-loop Bethe logarithms~\cite{PaJe2003}.
Our treatment relies on the identification of the imaginary
part of the energy shift in terms of a decay rate
of the reference state, as suggested
by Barbieri and Sucher in Ref.~\cite{BaSu1978}.
In Eq.~(\ref{NRQED_velocity}), $\delta \Gamma_i^{(1)}$ is a correction to the 
one-photon decay rate, whereas $\Gamma_i^{(2)}$ 
is the two-photon decay rate. The former is obtained by terms 
where the integration over $\omega_1$ or $\omega_2$ meets
a bound-state pole and generates an imaginary part, 
in the sense of Eq.~(4) of Ref.~\cite{Je2007}, but the 
other photon energy is integrated with a principal-value 
prescription. The latter term, $\Gamma_i^{(2)}$, is obtained by 
selecting exclusively the imaginary part generated by the singularities 
at $\omega_1 + \omega_2 = E_i - E_v$, where $E_v$ is a
virtual state contained in one of the propagators.
All expressions on the right-hand side of Eq.~(\ref{NRQED_velocity}) 
are manifestly of order $\alpha^2 (Z\alpha)^6 m$,
i.e.~$(Z\alpha)^6 R_\infty$ where $R_\infty$ is the
Rydberg constant.

The function $f_\epsilon$ reads as follows (with all infinitesimal 
imaginary parts duly taken into account),
\begin{align}
\label{f}
f_\epsilon(\omega_1, \omega_2) =&
\left< \phi_i \left| 
p^j \, \frac{1}{E - H - \omega_1 + {\rm i}\epsilon} \, p^k \, 
\frac{1}{E - H - \omega_1 - \omega_2 + {\rm i}\epsilon} \, p^j \, 
\frac{1}{E - H - \omega_2 + {\rm i}\epsilon} \, p^k  \right| \phi_i \right> 
\nonumber\\[1ex]
& + \frac{1}{2} \,
\left< \phi_i \left| 
p^j \, \frac{1}{E - H - \omega_1 + {\rm i}\epsilon} \, p^k \,
\frac{1}{E - H - \omega_1 - \omega_2 + {\rm i}\epsilon} \, p^k \, 
\frac{1}{E - H - \omega_1 + {\rm i}\epsilon} \, p^j  \right| \phi_i \right> 
\nonumber\\[1ex] 
& + \frac{1}{2} \,
\left< \phi_i \left| 
p^j \, \frac{1}{E - H - \omega_2 + {\rm i}\epsilon} \, p^k \, 
\frac{1}{E - H - \omega_1 - \omega_2 + {\rm i}\epsilon} \, p^k \, 
\frac{1}{E - H - \omega_2 + {\rm i}\epsilon} \, p^j  \right| \phi_i \right>
+ \dots
\end{align}
where the terms denoted by 
the ellipsis are given in Eq.~(3) of Ref.~\cite{Je2007},
being irrelevant for the current investigation, because the two-photon 
decay rate is generated exclusively by the poles where the sum
$\omega_1 + \omega_2$ of both photon energies is on resonance.
In a basis-set representation, the expression for the two-photon decay rate 
$\Gamma^{(2)}$ is thus found from the first three 
terms in Eq.~(\ref{f}) as~\cite{Je2007}
\begin{align}
\label{decay_velocity}
\Gamma^{(2)} =& \; \frac{4 \alpha^2}{9 \pi m^2} \,
{\rm Re}
\int\limits_0^{E_i - E_f}
{\rm d}\omega \, \omega \, (E_i - E_f - \omega) \, 
\left( \sum_v 
\left\{ 
\frac{\left< \phi_f \left| p^k \right| \phi_v \right> 
\left< \phi_v \left| p^j \right| \phi_i \right> }
  {E_i - E_v  - \omega + {\rm i} \epsilon} +
\frac{\left< \phi_f \left| p^k \right| \phi_v \right> 
\left< \phi_v \left| p^j \right| \phi_i \right> }
{E_f - E_v + \omega + {\rm i} \epsilon} \right\} \right) \,
\nonumber\\[2ex]
&  
\times \left( \sum_w 
\left\{ 
\frac{\left< \phi_i \left| p^k \right| \phi_w \right> 
\left< \phi_w \left| p^j \right| \phi_f \right> }
  {E_i - E_w  - \omega + {\rm i} \epsilon} +
\frac{\left< \phi_i \left| p^k \right| \phi_w \right> 
\left< \phi_w \left| p^j \right| \phi_f \right> }
{E_f - E_w + \omega + {\rm i} \epsilon} \right\} \right) \, \,,
\end{align}
where we use the summation convention for the 
Cartesian coordinates labeled by the indices 
$j \in \{1,2,3\}$ and $k \in \{1,2,3\}$.
The sum over $v$ contains all virtual states,
i.e.~over the entire bound and continuous spectrum.
We here imply a sum over the magnetic projections of the intermediate states,
and of the final state of the decay process, 
but an averaging over magnetic projections
of the initial state (since the decay rate does not 
depend on the magnetic projection of the initial state, 
one may alternatively choose any allowed value for the 
initial-state magnetic projection).

We now assume all initial and final, and virtual 
states to be given in terms of 
hydrogen wave functions in the standard representation 
(see, e.g.,~Ref.~\cite{LaLi1958}),
so that 
\begin{equation}
\label{standard_velocity}
\left< \phi_f \left| p^j \right| \phi_v \right> 
\left< \phi_v \left| p^j \right| \phi_i \right> =
\left< \phi_i \left| p^j \right| \phi_v \right> 
\left< \phi_v \left| p^j \right| \phi_f \right> \,.
\end{equation}
where the sum over $j$ is assumed. 
We then do the angular algebra~\cite{VaMoKh1988}.
For $nS \to 1S$ decays, one obtains a result~\cite{Je2007}
which reproduces the well-known expression obtained 
by G\"{o}ppert--Mayer in Ref.~\cite{GM1931}
for the particular case of $| \phi_i \rangle = | 2S \rangle$,
\begin{equation}
\label{decay_velocity_nS1S}
\Gamma^{(2)}_{nS} = \frac{4 \alpha^2}{27 \pi m^2} 
\lim_{\epsilon \to 0} 
{\rm Re} \!\!\!
\int\limits_0^{E_{nS} - E_{1S}}
{\rm d}\omega \, \omega \, (E_{nS} - E_{1S} - \omega) \, 
\left( \sum_{\nu}
\left\{ 
\frac{\left< 1S \left|\left| \vec{p} \right|\right| \nu P \right> 
\left< \nu P \left|\left| \vec{p} \right|\right| nS \right> }
  {E_{nS} - E_{\nu P}  - \omega + {\rm i} \epsilon} +
\frac{\left< 1S \left|\left| \vec{p} \right|\right| \nu P \right> 
\left< \nu P \left|\left| \vec{p} \right|\right| nS \right> }
{E_{1S} - E_{\nu P} + \omega + {\rm i} \epsilon} \right\} \right)^2 ,
\end{equation}
where we use the definition of the reduced matrix elements according 
to Ref.~\cite{VaMoKh1988}.
Virtual $P$ states are also relevant for the 
decay $nD \to 1S$ decays, but the well-known 
prefactor is different~\cite{CrTaSaCh1986}, and the result is 
\begin{equation}
\label{decay_velocity_nD1S}
\Gamma^{(2)}_{nD} = \frac{4 \alpha^2}{135 \pi m^2} 
\lim_{\epsilon \to 0} 
{\rm Re} \!\!\!\!\!\!
\int\limits_0^{E_{nD} - E_{1S}} \!\!\!\!\!\!
{\rm d}\omega \, \omega \, (E_{nD} - E_{1S} - \omega) \, 
\left( \sum_{\nu}
\left\{ 
\frac{\left< 1S \left|\left| \vec{p} \right|\right| \nu P \right> 
\left< \nu P \left|\left| \vec{p} \right|\right|  nD \right> }
  {E_{nD} - E_{\nu P}  - \omega + {\rm i} \epsilon} +
\frac{\left< 1S \left|\left| \vec{p} \right|\right|  \nu P \right> 
\left< \nu P \left|\left| \vec{p} \right|\right|  nD \right> }
{E_{1S} - E_{\nu P} + \omega + {\rm i} \epsilon} \right\} \right)^2,
\end{equation}
where for completeness we note that the reduced matrix element 
for $P \to D$ transitions differs from the ``radial'' component of the 
matrix element by a factor $\sqrt{2}$.

In the length gauge, the atom-field interaction is 
given by 
\begin{equation}
H_I = - e\, \vec{E}\cdot \vec{r} \,,
\end{equation}
where $\vec{E}$ is the quantized electric-field operator.
The length-gauge two-photon self-energy is obtained 
by straightforward fourth-order perturbation theory as
\begin{equation}
\label{NRQED_length}
\Delta E^{(2)}_i = \lim_{\epsilon \to 0}
\left( \frac{2 \alpha}{3 \pi m^2} \right)^2
\int_0^{\Lambda_1} d\omega_1 \, \omega_1^3 
\int_0^{\Lambda_2} d\omega_2 \, \omega_2^3 \, 
g_\epsilon(\omega_1, \omega_2) = 
{\rm Re} \, \Delta E^{(2)}_i 
- {\rm i} \, \frac{\delta \Gamma_i^{(1)}}{2}
- {\rm i} \, \frac{\Gamma_i^{(2)}}{2}\,.
\end{equation}
We observe the factor $\omega_1^3 \, \omega_2^3$, which is 
characteristic of the length-gauge formulation.
The absence of the seagull term as opposed to the velocity gauge
leads to a somewhat simplified expression,
\begin{align}
g_\epsilon(\omega_1, \omega_2) =&
\left< \phi_i \left|
x^j \, \frac{1}{E - H - \omega_1 + {\rm i}\epsilon} \, x^k \,
\frac{1}{E - H - \omega_1 - \omega_2 + {\rm i}\epsilon} \, x^j \,
\frac{1}{E - H - \omega_2 + {\rm i}\epsilon} \, x^k  \right| \phi_i \right>
\nonumber\\[1ex]
& + \frac{1}{2} \,
\left< \phi_i \left|
x^j \, \frac{1}{E - H - \omega_1 + {\rm i}\epsilon} \, x^k \,
\frac{1}{E - H - \omega_1 - \omega_2 + {\rm i}\epsilon} \, x^k \,
\frac{1}{E - H - \omega_1 + {\rm i}\epsilon} \, x^j  \right| \phi_i \right>
\nonumber\\[1ex]
& + \frac{1}{2} \,
\left< \phi_i \left|
x^j \, \frac{1}{E - H - \omega_2 + {\rm i}\epsilon} \, x^k \,
\frac{1}{E - H - \omega_1 - \omega_2 + {\rm i}\epsilon} \, x^k \,
\frac{1}{E - H - \omega_2 + {\rm i}\epsilon} \, x^j  \right| \phi_i \right>
\nonumber\\[1ex]
& +
\left< \phi_i \left|
x^j \, \frac{1}{E - H - \omega_1 + {\rm i}\epsilon} \, x^j \,
\left( \frac{1}{E - H} \right)' \, x^k \,
\frac{1}{E - H - \omega_2 + {\rm i}\epsilon} \, x^k  \right| \phi_i \right>
\nonumber\\[1ex]
& - \frac{1}{2} \,
\left< \phi_i \left|
x^j \, \frac{1}{E - H - \omega_1 + {\rm i}\epsilon} \, x^j
\right| \phi_i \right> \,
\left< \phi_i \left|
x^k \, \left( \frac{1}{E - H - \omega_2 + {\rm i}\epsilon} \right)^2 \,
x^k  \right| \phi_i \right>
\nonumber\\[1ex]
& - \frac{1}{2} \,
\left< \phi_i \left|
x^j \, \frac{1}{E - H - \omega_2 + {\rm i}\epsilon} \,
x^j  \right| \phi_i \right> \,
\left< \phi_i \left|
x^k \, \left( \frac{1}{E - H - \omega_1 + {\rm i}\epsilon} \right)^2 \,
x^k  \right| \phi_i \right>\,.
\end{align}
In contrast to Eq.~(\ref{f}), the momentum operators are replaced by 
position operators. In analogy to Eq.~(\ref{f}), 
only the first three terms are relevant for the 
two-photon decay rate.
Using a basis-set representation, the expression for the two-photon decay rate 
derived in the length gauge thus reads
\begin{align}
\label{decay_length}
\Gamma^{(2)} =& \; \frac{4 \alpha^2}{9 \pi m^2} \,
{\rm Re}
\int\limits_0^{E_i - E_f}
{\rm d}\omega \, \omega^3 \, (E_i - E_f - \omega)^3 \, 
\left( \sum_v 
\left\{ 
\frac{\left< \phi_f \left| x^k \right| \phi_v \right> 
\left< \phi_v \left| x^j \right| \phi_i \right> }
  {E_i - E_v  - \omega + {\rm i} \epsilon} +
\frac{\left< \phi_f \left| x^k \right| \phi_v \right> 
\left< \phi_v \left| x^j \right| \phi_i \right> }
{E_f - E_v + \omega + {\rm i} \epsilon} \right\} \right) \,
\nonumber\\[2ex]
&  
\times \left( \sum_w 
\left\{ 
\frac{\left< \phi_i \left| x^k \right| \phi_w \right> 
\left< \phi_w \left| x^j \right| \phi_f \right> }
  {E_i - E_w  - \omega + {\rm i} \epsilon} +
\frac{\left< \phi_i \left| x^k \right| \phi_w \right> 
\left< \phi_w \left| x^j \right| \phi_f \right> }
{E_f - E_w + \omega + {\rm i} \epsilon} \right\} \right) \, \,,
\end{align}
where the sum over $v$ contains all virtual states,
Using the identity
\begin{align}
\label{oak}
& \sum_v 
\left\{ 
\frac{\left< \phi_f \left| p^k \right| \phi_v \right> 
\left< \phi_v \left| p^j \right| \phi_i \right> }
  {E_i - E_v  - \omega + {\rm i} \epsilon} +
\frac{\left< \phi_f \left| p^k \right| \phi_v \right> 
\left< \phi_v \left| p^j \right| \phi_i \right> }
{E_f - E_v + \omega + {\rm i} \epsilon} \right\} 
\nonumber\\[2ex]
& \quad =
\omega \, (E_i - E_f - \omega) \,
\sum_v 
\left\{ 
\frac{\left< \phi_f \left| x^k \right| \phi_v \right> 
\left< \phi_v \left| x^j \right| \phi_i \right> }
  {E_i - E_v  - \omega + {\rm i} \epsilon} +
\frac{\left< \phi_f \left| x^k \right| \phi_v \right> 
\left< \phi_v \left| x^j \right| \phi_i \right> }
{E_f - E_v + \omega + {\rm i} \epsilon} \right\} 
\end{align}
it is easy to show the equivalence of the two expressions for the 
two-photon decay rate given in Eqs.~(\ref{NRQED_velocity}) 
and~(\ref{NRQED_length}). Note that this equivalence can be shown 
easily using the commutator relation $p^i = {\rm i} \, [H, x^j]$, but 
it holds only if the sum over $v$ extends over the complete spectrum.

Assuming hydrogen wave functions in the standard representation,
we have that in analogy to Eq.~(\ref{standard_velocity}), 
\begin{equation}
\label{standard_length}
\left< \phi_f \left| x^j \right| \phi_v \right> 
\left< \phi_v \left| x^j \right| \phi_i \right> =
\left< \phi_i \left| x^j \right| \phi_v \right> 
\left< \phi_v \left| x^j \right| \phi_f \right> \,.
\end{equation}
After angular algebra, one obtains for the 
decay $nS \to 1S$,
\begin{equation}
\label{decay_length_nS1S}
\Gamma^{(2)}_{nS} = \frac{4 \alpha^2}{27 \pi m^2} \,
\lim_{\epsilon \to 0} \, 
{\rm Re} \!\!\!\!\!\!
\int\limits_0^{E_{nS} - E_{1S}} \!\!\!\!\!\!
{\rm d}\omega \, \omega^3 \, (E_{nS} - E_{1S} - \omega)^3 \, 
\left( \sum_{\nu}
\left\{ 
\frac{\left< 1S \left|\left| \vec{x} \right|\right| \nu P  \right> 
\left< \nu P \left|\left| \vec{x} \right|\right| nS \right> }
  {E_{nS} - E_{\nu P}  - \omega + {\rm i} \epsilon} +
\frac{\left< 1S \left|\left| \vec{x} \right|\right| \nu P  \right> 
\left< \nu P \left|\left| \vec{x} \right|\right| nS \right> }
{E_{1S} - E_{\nu P} + \omega + {\rm i} \epsilon} \right\} \right)^2 ,
\end{equation}
whereas for $nD \to 1S$ decays, 
\begin{equation}
\label{decay_length_nD1S}
\Gamma^{(2)}_{nD} = \frac{4 \alpha^2}{135 \pi m^2} 
\lim_{\epsilon \to 0} 
{\rm Re} \!\!\!\!\!\!
\int\limits_0^{E_i - E_f} \!\!\!\!\!\!
{\rm d}\omega \, \omega^3 \, (E_{nD} - E_{1S} - \omega)^3 
\left( \sum_{\nu}
\left\{ 
\frac{\left< 1S \left|\left| \vec{x} \right|\right| \nu P \right> 
\left< \nu P \left|\left| \vec{x} \right|\right| nD \right> }
  {E_{nD} - E_{\nu P}  - \omega + {\rm i} \epsilon} +
\frac{\left< 1S \left|\left| \vec{x} \right|\right| \nu P  \right> 
\left< \nu P \left|\left| \vec{x} \right|\right| nD \right> }
{E_{1S} - E_{\nu P} + \omega + {\rm i} \epsilon} \right\} \right)^2, 
\end{equation}
again in complete analogy to Eqs.~(\ref{decay_velocity_nS1S})
and~(\ref{decay_velocity_nD1S}), respectively.

%
%
\section{Numerical Results} 
\label{numerical_results}

We here focus on the $nS \to 1S$ and $nD \to 1S$ decays, 
as indicated in Eqs.~(\ref{decay_length_nS1S}) and~(\ref{decay_length_nD1S}),
respectively.
Decays to the ground state have the highest rate for 
both one-photon~\cite{BeSa1957} as well as two-photon processes
and are therefore of special interest.
Due to the infinitesimal imaginary parts explicitly indicated in 
Eqs.~(\ref{decay_length_nS1S}) and~(\ref{decay_length_nD1S}), 
we can extend the sum over intermediate, 
virtual states over the entire hydrogenic spectrum, 
including those $P$ states which have a lower energy than the 
reference state. We recall here that the double poles at intermediate
resonances are naturally treated using the formula~\cite{Je2007}
\begin{equation}
\label{model}
\lim_{\epsilon \to 0}  {\rm Re}
\int_0^1 {\rm d}\omega \, 
\left( \frac{1}{a - \omega + {\rm i}\epsilon} \right)^2 
= \frac{1}{a(a-1)} \,.
\end{equation}
Simple poles are treated using the well-known Dirac prescription,
and the principal-value integration then yields the real part of the 
integrals. Numerical results can be obtained by expressing the 
matrix elements with the propagators in terms in hypergeometric functions,
following Refs.~\cite{GaCo1970,Pa1993}.
Final values are indicated in Table~\ref{results}.

\begin{table}[thb]
\begin{center}
\begin{minipage}{13cm}
\begin{center}
\caption{\label{results} 
Numerical results for the decay rates $nS \to 1S$ and 
$nD \to 1S$ for hydrogen. The rates 
scale with $Z^6$ for a hydrogenlike ions with nuclear charge 
number $Z$. Units are
inverse seconds. To obtain the decay rate in Hertz, one needs to divide
by a factor of $2\pi$. We here supplement the results given in 
Ref.~\cite{Je2007} by some values for higher excited $S$ states
and we also indicated results for $nD \to 1S$, which were 
not treated in Ref.~\cite{Je2007}.}
\begin{tabular}{c@{\hspace{0.8cm}}c@{\hspace{0.8cm}}c@{\hspace{0.8cm}}%
c}
\hline
\hline
\rule[-2mm]{0mm}{6mm}
&
$|\phi_f \rangle = | 1{\rm S} \rangle$ &
&
$|\phi_f \rangle = | 1{\rm S} \rangle$ \\
\hline
\rule[-2mm]{0mm}{6mm}
$|\phi_i \rangle = | 2{\rm S} \rangle$ & $8.229\,352$ & 
 & \\
\rule[-2mm]{0mm}{6mm}
$|\phi_i \rangle = | 3{\rm S} \rangle$ & $2.082\,853$ & 
$|\phi_i \rangle = | 3{\rm D} \rangle$ & $1.042\,896$ \\
\rule[-2mm]{0mm}{6mm}
$|\phi_i \rangle = | 4{\rm S} \rangle$ & $0.698\,897$ &
$|\phi_i \rangle = | 4{\rm D} \rangle$ & $0.598\,798$ \\
\rule[-2mm]{0mm}{6mm}
$|\phi_i \rangle = | 5{\rm S} \rangle$ & $0.287\,110$ &
$|\phi_i \rangle = | 5{\rm D} \rangle$ & $0.340\,883$ \\
\rule[-2mm]{0mm}{6mm}
$|\phi_i \rangle = | 6{\rm S} \rangle$ & $0.135\,935$ &
$|\phi_i \rangle = | 6{\rm D} \rangle$ & $0.206\,523$ \\
\rule[-2mm]{0mm}{6mm}
$|\phi_i \rangle = | 7{\rm S} \rangle$ & $0.071\,402$ &
$|\phi_i \rangle = | 7{\rm D} \rangle$ & $0.132\,928$ \\
\rule[-2mm]{0mm}{6mm}
$|\phi_i \rangle = | 8{\rm S} \rangle$ & $0.040\,587$ &
$|\phi_i \rangle = | 8{\rm D} \rangle$ & $0.090\,016$ \\
\hline
\hline
\end{tabular}
\end{center}
\end{minipage}
\end{center}
\end{table}

The one-loop as well as the two-loop self-energy
shifts of hydrogenic states are well known to follow 
scaling laws of the form of inverse powers 
of the principal quantum number $n$, as analyzed in Ref.~\cite{JeEtAl2003}.
The two-photon decay rate is the imaginary part of this 
energy shift and is thus expected to follow an analogous 
trend with the principal quantum numbers.
Analyzing the data in Table~\ref{results},
we find that the $nD \to 1S$ state results appear to 
follow the asymptotic behaviour (expressed in inverse 
seconds)
\begin{equation}
\label{asymp_nD}
\Gamma^{(2)}_{nD} = \frac{49(2)}{n^3} \, Z^6 \, s^{-1} \,, \qquad
n \to \infty \,,
\end{equation}
whereas for $nS \to 1S$ decay, a fractional power apparently leads to a 
more satisfactory representation of the data,
\begin{equation}
\label{asymp_nS}
\Gamma^{(2)}_{nS} = \frac{330(20)}{n^{4.3}} \, Z^6 \, s^{-1}  \,, \qquad
n \to \infty\,.
\end{equation}
The results indicated in Table~\ref{results}
are consistent with a decrease of the two-photon decay rate 
with increasing $n$.

%
%
\section{Discussion and Comparison}
\label{discussion}

When comparing to the existing literature, it is useful, 
first of all, to note the calculations~\cite{Fl1984,FlScMi1988,ChSu2007},
which are apparently based on second-order perturbation theory 
for the two-photon transition amplitude. 
As a consequence, they present singularities when 
the energy of one of the photons reaches a level situated
between the initial and final states, and no procedure is 
given in the cited references if one does not go beyond second order.
When evaluating differential transition rates 
(Refs.~\cite{Fl1984,TuYeSaCh1984,FlPaSt1987}), the absence of the 
infinitesimal imaginary part does not matter, and the numerical 
results in the velocity gauge~\cite{Fl1984,FlPaSt1987} and in the 
length gauge~\cite{TuYeSaCh1984} fully agree.
The problem arises when one tries to evaluate the total 
decay rate, as the existing singularities are not integrable.
Although in Ref.~\cite{CrTaSaCh1986} fourth-order perturbation
theory was used, a consistent answer does not appear to have been found.

It appears that in general, two approaches have been used so far in the 
literature in order to deal with the problematic double poles 
for the photon energy integrations: (i) the explicit removal
of particular states from the sum over virtual states, and 
(ii) the inclusion of a width for the intermediate, virtual 
states. 

Let us begin the discussion with the removal of states.
Indeed, Chluba and Sunyaev~\cite{ChSu2007}, 
Florescu {\em et al.}~\cite{FlScMi1988} as well as
Cresser {\em et al.}~\cite{CrTaSaCh1986} 
have used different formulas than those used here,
in order to evaluate the two-photon 
decay rates. In particular,
they use instead of Eq.~(\ref{decay_length_nS1S}) 
the following formula for $nS \to 1S$ decays,
\begin{equation}
\label{decay_length_nS1S_subtracted}
\gamma^{(2)}_{nS} = \frac{4 \alpha^2}{27 \pi m^2} \,
\int\limits_0^{E_{nS} - E_{1S}}
{\rm d}\omega \, \omega^3 \, (E_{nS} - E_{1S} - \omega)^3 \, 
\left| \sum_{\nu \geq N}
\left\{ 
\frac{\left< 1S \left| \left| \vec{x} \right| \right| \nu P  \right> 
\left< \nu P \left| \left| \vec{x} \right| \right| nS \right> }
  {E_{nS} - E_{\nu P} - \omega} +
\frac{\left< 1S \left| \left| \vec{x} \right| \right| \nu P  \right> 
\left< \nu P \left| \left| \vec{x} \right| \right| nS \right> }
{E_{1S} - E_{\nu P} + \omega} \right\} \right|^2 \,,
\end{equation}
where $N = n$ (Chluba and Sunyaev, Ref.~\cite{ChSu2007}) or
or $N = n+1$ (Florescu {\em et al.}~\cite{Fl1984,FlScMi1988} 
and Cresser {\em et al.}, Ref.~\cite{CrTaSaCh1986}), and 
the notation $\nu \geq N$ of course means that one should sum over the 
discrete spectrum for all virtual states with principal 
quantum numbers as indicated, and of course integrate over the 
entire continuum spectrum in addition.
For $nD \to 1S$ decays, the cited authors use
\begin{equation}
\label{decay_length_nD1S_subtracted}
\gamma^{(2)}_{nD} = \frac{4 \alpha^2}{135 \pi m^2} \,
\int\limits_0^{E_{nD} - E_{1S}}
{\rm d}\omega \, \omega^3 \, (E_{nD} - E_{1S} - \omega)^3 \, 
\left| \sum_{\nu \geq N}
\left\{ 
\frac{\left< 1S \left| \left| \vec{x} \right| \right| \nu P \right> 
\left< \nu P \left| \left| \vec{x} \right| \right| nD \right> }
  {E_{nD} - E_{\nu P}  - \omega} +
\frac{\left< 1S \left| \left| \vec{x} \right| \right| \nu P  \right> 
\left< \nu P \left| \left| \vec{x} \right| \right| nD \right> }
  {E_{1S} - E_{\nu P} + \omega} \right\} \right|^2 \,,
\end{equation}
with the same proposed values for $N$.
In this case, because the problematic 
virtual states of lower energy than the initial state 
$|\phi_i\rangle$ have been explicitly removed from the sum over 
virtual states, there are no more singularities infinitesimally 
displaced from the integration contours present, and there is
therefore no need for any infinitesimal imaginary
part ${\rm i}\, \epsilon$ in the propagator denominators.
Furthermore, $| \cdot |^2$ is equivalent to $( \cdot )^2$
provided our assumption formulated in 
Eq.~(\ref{standard_velocity}) holds.
The corresponding velocity-gauge expressions,
\begin{equation}
\label{decay_velocity_nS1S_subtracted}
\eta^{(2)}_{nS} = \frac{4 \alpha^2}{27 \pi m^2} \,
\int\limits_0^{E_{nS} - E_{1S}}
{\rm d}\omega \, \omega \, (E_{nS} - E_{1S} - \omega) \, 
\left| \sum_{\nu \geq N}
\left\{ 
\frac{\left< 1S \left| \left| \vec{p} \right| \right| \nu P  \right> 
\left< \nu P \left| \left| \vec{p} \right| \right| nS \right> }
  {E_{nS} - E_{\nu P}  - \omega} +
\frac{\left< 1S  \left| \left| \vec{p} \right| \right| \nu P  \right> 
\left< \nu P  \left| \left| \vec{p} \right| \right| nS \right> }
  {E_{1S} - E_{\nu P} + \omega} \right\} \right|^2 \,,
\end{equation}
and
\begin{equation}
\label{decay_velocity_nD1S_subtracted}
\eta^{(2)}_{nD} = \frac{4 \alpha^2}{135 \pi m^2} \,
\int\limits_0^{E_{nD} - E_{1S}}
{\rm d}\omega \, \omega \, (E_{nD} - E_{1S} - \omega) \, 
\left| \sum_{\nu \geq N}
\left\{ 
\frac{\left< 1S \left| \left| \vec{p} \right| \right| \nu P \right> 
\left< \nu P \left| \left| \vec{p} \right| \right| nD \right> }
  {E_{nD} - E_{\nu P}  - \omega} +
\frac{\left< 1S \left| \left| \vec{p} \right| \right| \nu P  \right> 
\left< \nu P \left| \left| \vec{p} \right| \right| nD \right> }
  {E_{1S} - E_{\nu P} + \omega} \right\} \right|^2 \,,
\end{equation}
are not equivalent to the length-gauge expressions
in Eqs.~(\ref{decay_length_nS1S_subtracted})
and~(\ref{decay_length_nD1S_subtracted}), because the 
relation (\ref{oak}) breaks down if the sum over $v$ 
does not extend over the entire hydrogen spectrum.
The explicit removal of the ``problematic'' virtual states
from the propagators avoids the necessity of
indicating the infinitesimal 
imaginary terms in the propagator denominators,
but the removal operation
leads to different expressions in the length and the 
velocity gauges and is thus not gauge invariant.

To illustrate this finding by a numerical example,
we observe that we can reproduce the value of
$\gamma^{(2)}_{3D} = 0.131 \, 813 \, {\rm s}^{-1}$
for the decay $3D \to 1S$ with $N = n+1$ using the 
length-gauge expression (\ref{decay_length_nD1S_subtracted}), 
in agreement with Eq.~(20) of Ref.~\cite{CrTaSaCh1986}. 
However, the velocity gauge expression (\ref{decay_velocity_nD1S_subtracted})
gives a different result, namely
$\eta^{(2)}_{3D} = 0.439\, 368\, {\rm s}^{-1}$.
These two results have to be contrasted with the 
gauge-invariant result of $\Gamma^{(2)}_{3D} = 1.042\,896\,\, {\rm s}^{-1}$,
indicated in Table~\ref{results}.
For the decay $3S \to 1S$, the values are 
$\gamma^{(2)}_{3S} = 8.225\,796 \, {\rm s}^{-1}$
in agreement with Eq.~(19) of Ref.~\cite{CrTaSaCh1986},
and the velocity-gauge result with $2P$ and $3P$ virtual states removed
is $\eta^{(2)}_{3S} = 6.192\,881 \, {\rm s}^{-1}$, whereas the 
gauge-invariant result with the full hydrogenic spectrum of 
virtual states reads
$\Gamma^{(2)}_{3S} = 2.082\,853\, {\rm s}^{-1}$
(see Table~\ref{results}).
It is interesting to observe that 
$\Gamma^{(2)}_{3D} > \gamma^{(2)}_{3D}$, but
$\Gamma^{(2)}_{3S} < \gamma^{(2)}_{3S}$.

Let us now turn our attention to the inclusion of a decay width 
for the intermediate states.
Indeed, the authors of Refs.~\cite{Fl1984,FlScMi1988,CrTaSaCh1986,ChSu2007}
arrive at the expressions~(\ref{decay_velocity_nS1S_subtracted})
and~(\ref{decay_velocity_nD1S_subtracted}) after analyzing 
the expression (for illustrative purposes we restrict ourselves
here to the $nS \to 1S$ decay)
\begin{equation}
\label{expr}
\frac{4 \alpha^2}{27 \pi m^2} \,
\int\limits_0^{E_{nS} - E_{1S}}
{\rm d}\omega \, \omega^3 \, (E_{nS} - E_{1S} - \omega)^3 \, 
\left| \sum_{\nu}
\left\{ 
\frac{\left< 1S  \left| \left| \vec{x} \right| \right| \nu P  \right> 
\left< \nu P  \left| \left| \vec{x} \right| \right| nS \right> }
  {E_{nS} - E_{\nu P}  - \omega + 
   \frac{\rm i}{2} \Gamma^{(1)}_v} +
\frac{\left< 1S  \left| \left| \vec{x} \right| \right| \nu P  \right> 
\left< \nu P  \left| \left| \vec{x} \right| \right| nS \right> }
  {E_{1S} - E_{\nu P} + \omega + 
   \frac{\rm i}{2} \Gamma^{(1)}_v} \right\} \right|^2 \,.
\end{equation}
Let us consider $3S$ state as an example.
The only ``problematic'' virtual state is the $2P$ state ($\nu = 2$),
and using the formula
\begin{equation}
\label{expr2}
\int_0^1 {\rm d}\omega \, 
\left| \frac{1}{a - \omega + {\rm i}\Gamma} \right|^2 = 
\frac{\pi}{\Gamma} + \frac{1}{a(a-1)} + 
{\cal O}(\Gamma^2) \,,
\end{equation}
it is possible to show, that the term with $\nu = 2$ 
in the expression~(\ref{expr}) gives rise to a
contribution which is equivalent to the 
one-photon decay rate $3S \to 2P$,
and this decay rate is just the total one-photon 
decay rate of the $3S$ state,
and it is equal to the
imaginary part of the one-loop self-energy of the $3S$ state
(in the dipole approximation).
The authors of Refs.~\cite{Fl1984,FlScMi1988,CrTaSaCh1986,ChSu2007}
thus conclude that this term should be interpreted 
as the one-photon decay rate of the $3S$ state,
which has got nothing to do with the two-photon decay process,
and this observation appears to be the basis for their 
removal of the $2P$ state from the sum over virtual 
states (see also the analysis in footnote~4 of 
Ref.~\cite{CrTaSaCh1986}).

Despite the appealing aspects of the removal 
operation, it is unfortunately not gauge invariant, 
as shown above, and we would like to point out two more
aspects that merit a discussion.
First and foremost, the discussion in 
footnote~4 of Ref.~\cite{CrTaSaCh1986} shows that 
the expression~(\ref{expr}) gives rise to a one-photon 
decay rate, effectively mixing the two-loop self-energy
with the one-photon self-energy (according to
the interpretation of a decay rate as an imaginary
part of an energy shift). If one would take 
the expression~(\ref{expr}) literally, then one should be careful
to avoid a double counting of the 
one-photon decay rate, which is already contained in the 
one-loop self-energy and should not be obtained once more
from the imaginary part of the two-loop self-energy.
Cascade contributions are discussed in more detail in 
Sec.~\ref{cascades} below.

The second aspect is observed when the 
analysis in footnote~4 of Ref.~\cite{CrTaSaCh1986} is
generalized to the $4S \to 1S$ decay. In that case,
two cascades are possible, namely~$4S \to 3P \to 1S$ and
$4S \to 2P \to 1S$. As an easy generalization 
of the analysis in footnote~4 of Ref.~\cite{CrTaSaCh1986} shows,
the full one-photon decay rate of the $4S$ state
is obtained from the expression (\ref{expr}) 
only if the virtual $2P$ and $3P$ are endowed with their
{\em partial} decay rates to the $1S$ ground state, i.e.~the 
$3P$ decay rate should be inserted into the 
propagator denominators as the {\em partial}
decay rate $3P \to 1S$, excluding the decay process
$3P \to 2S$. If one generalizes these considerations further,
namely to a general decay $nS \to 1S$, then this 
would imply that one should use different
decay rates $\Gamma^{(1)}_v$ in Eq.~(\ref{expr}) to regularize the 
divergence in $1/\Gamma^{(1)}_v$ in Eq.~(\ref{expr2}),
adjusting them according to the decay process under study. 
That prescription would be highly counterintuitive as the 
virtual states should somehow ``know'' about properties of the 
initial and final states of the decay process. The ensuing questions
have already been noticed by Chluba and Sunyaev~\cite{ChSu2007}.

Let us conclude this section with a remark 
on the asymptotics (\ref{asymp_nD}) and (\ref{asymp_nS}),
which permit an extrapolation of our results to 
Rydberg states with high principal quantum numbers.
Some investigations, including Ref.~\cite{ChSu2007},
lead to results for the two-photon decay rates of higher 
excited state which exhibit a linear increase with $n$
instead of a decrease with at least $n^{-3}$,
as indicated in Eqs.~(\ref{asymp_nD}) and (\ref{asymp_nS}).
It is well-known that the one-photon rates decrease
approximately with $n^{-3}$ (see Ref.~\cite{BeSa1957}).
If the two-photon rates would indeed increase linearly,
then there would be a relative factor $n^4$ with which
two-photon rates would grow in comparison to one-photon rates
as the principal quantum number of a state increases.
If we take into account the relative scaling factor
of $Z^2 \alpha^3/\pi$ by which two-photon rates are suppressed
with respect to one-photon rates, then we would have to conclude that
the two-photon rates overtake the one-photon rates
for states with a comparatively low principal quantum
number of $n \approx 50/\sqrt{Z}$ in a hydrogenlike ion
with nuclear charge number $Z$.
For our results as indicated in Table~\ref{results}, 
the two-photon rates are suppressed
with respect to one-photon rates by a relative
factor  $Z^2 \alpha^3/\pi$ for all hydrogenic states,
because the scaling with $n$ is obtained to be approximately
the same for the one- as well as the two-photon rates,
and the natural hierarchy of the likelihood of one- and two-photon 
events is preserved for all states.

%
%
\section{Extraction of the Cascade Contribution}
\label{cascades}

As in Sec.~\ref{discussion}, let us focus on a particular
example whose generalization is obvious,
namely (this time) the $3S \to 1S$ decays, for which the 
cascade $3S \to 2P \to 1S$ needs to be addressed.
Let us go back
once more to Eq.~(\ref{decay_length_nS1S}),
\begin{equation}
\label{decay_length_3S1S}
\Gamma^{(2)}_{3S} = \frac{4 \alpha^2}{27 \pi m^2} \,
\lim_{\epsilon \to 0} \, 
{\rm Re} \!\!\!\!\!\!
\int\limits_0^{E_{3S} - E_{1S}} \!\!\!\!\!\!
{\rm d}\omega \, \omega^3 \, (E_{3S} - E_{1S} - \omega)^3 \, 
\left( \sum_{\nu}
\left\{ 
\frac{\left< 1S  \left| \left| \vec{x} \right| \right| \nu P  \right> 
\left< \nu P  \left| \left| \vec{x} \right| \right| 3S \right> }
  {(E_{3S} - E_{\nu P}  - \omega + {\rm i} \epsilon} +
\frac{\left< 1S  \left| \left| \vec{x} \right| \right| \nu P  \right> 
\left< \nu P  \left| \left| \vec{x} \right| \right| 3S \right> }
  {(E_{1S} - E_{\nu P} + \omega + {\rm i} \epsilon} \right\} \right)^2,
\end{equation}
and adopt the cumbersome, but absolutely unique 
notation P.V.~for the principal value part of the distribution.
If we use the formula
\begin{equation}
\label{formula}
\frac{1}{x + {\rm i} \, \epsilon} = 
({\rm P.V.}) \frac{1}{x} - 
{\rm i}\pi \, \delta(x) 
\end{equation}
for all propagator denominators in Eq.~(\ref{decay_length_3S1S})
and extract only the contribution due to the 
delta functions, then the only contributing virtual 
state is the $2P$ state. Because there is a product
of two terms both of which become singular, we cannot avoid
to obtain the square of the delta function,
\begin{equation}
\delta^2(\omega - E_{3S} - E_{2P}) = 
\delta(0) \, \delta(\omega - E_{3S} - E_{2P}) = 
\frac{T}{2 \pi} \, \delta(\omega - E_{3S} - E_{2P}) \,,
\end{equation}
and a further term proportional to
$\delta(\omega - E_{2P} + E_{1S}) \, T/(2\pi)$.
Here, $T$ is the (long) observation time proportional
to $\delta(0)$ in energy space (see, e.g. Ref.~\cite{Ri1972}).
The sum of the terms proportional to $\delta(0)$ reads
\begin{equation}
\label{cascade1_3S1S}
C^{(2)}_{3S} = -T \, 
\Gamma^{(1)}_{3S \to 2P} \,
\Gamma^{(1)}_{2P \to 1S} \,,
\qquad
\int {\rm d} T \, C^{(2)}_{3S} = 
-\frac{T^2}{2} \, \Gamma^{(1)}_{3S \to 2P} \,
\Gamma^{(1)}_{2P \to 1S} \,,
\end{equation}
where we introduce an obvious notation for the partial one-photon 
rates $\Gamma^{(1)}_{3S \to 2P}$ and $\Gamma^{(1)}_{2P \to 1S}$.
Note, in particular, that the resulting 
expression for $C^{(2)}_{3S}$ is gauge invariant.

Our result (\ref{cascade1_3S1S}) has just the right form 
to describe the cascade decay, except for the ``wrong'' sign. For the term 
to contribute to the decay of the $3S$ state, it should be positive, but it 
turns out to be negative. Let us defer a discussion of this issue
and instead consider the extraction of the cascade contribution from the 
expression 
\begin{equation}
\label{decay_length_3S1S_modulus}
{\tilde \Gamma}^{(2)}_{3S} = \frac{4 \alpha^2}{27 \pi m^2} \,
\lim_{\epsilon \to 0} \, 
{\rm Re} \!\!\!\!\!\!
\int\limits_0^{E_{3S} - E_{1S}}  \!\!\!\!\!\!
{\rm d}\omega \, \omega^3 \, (E_{3S} - E_{1S} - \omega)^3 \, 
\prod_\pm \sum_{\nu}
\left\{ 
\frac{\left< 1S  \left| \left| \vec{x} \right| \right| \nu P  \right> 
\left< \nu P  \left| \left| \vec{x} \right| \right| 3S \right> }
  {E_{3S} - E_{\nu P}  - \omega \pm {\rm i} \epsilon} +
\frac{\left< 1S \left| \left| \vec{x} \right| \right| \nu P  \right> 
\left< \nu P \left| \left| \vec{x} \right| \right| 3S \right> }
  {E_{1S} - E_{\nu P}  + \omega \pm {\rm i} \epsilon} \right\}  \,,
\end{equation}
where $\prod_\pm$ means the product of two terms, with either sign.
The product over the two terms with $\pm {\rm i} \epsilon$
is of course equivalent to the square of the modulus of the 
two terms in the integrand, in analogy to Eq.~(\ref{expr}).
If we now use (\ref{formula}), then we obtain
\begin{equation}
\label{cascade2_3S1S}
{\tilde C}^{(2)}_{3S} = T \, 
\Gamma^{(1)}_{3S \to 2P} \,
\Gamma^{(1)}_{2P \to 1S} \,,
\qquad
\int {\rm d} T \, {\tilde C}^{(2)}_{3S} = 
\frac{T^2}{2} \, \Gamma^{(1)}_{3S \to 2P} \,
\Gamma^{(1)}_{2P \to 1S}\,.
\end{equation}
This result has the ``right'' sign, and it has, for large $T$, the 
right temporal dependence for a cascade process.

In order to resolve the paradox, one first should notice that both signs
found in Eqs.~(\ref{cascade1_3S1S}) and~(\ref{cascade2_3S1S})
actually have a valid interpretation. The two-loop self-energy contains 
both radiative corrections to the one-photon decay as well as the 
full two-photon decay amplitude. The radiative corrections to one-photon
decay are obtained by ``cutting'' appropriate internal lines in the 
diagrams, and indeed, we can rederive the first three 
radiative corrections to one-photon as given in Eq.~(27) 
of Ref.~\cite{SaPaCh2004} by considering resonant intermediate states
in the ``outer'' electron propagators in the terms 
of Eq.~(\ref{f}). (The remaining terms used in Eq.~(27)
of Ref.~\cite{SaPaCh2004} follow from standard third-order 
perturbation theory.) The magnitude of the radiative corrections to 
one-photon decay is decreased by the possibility of cascade processes, 
due to the virtual-to-real conversion of the photons 
appearing in the integrals for the radiative corrections to the 
one photon decay at resonance, 
and this decrease is consistent with the 
sign of the right-hand side of Eq.~(\ref{cascade1_3S1S}). 
On the other hand, the two-photon decay amplitude should be increased
by the cascade processes, and this increase is consistent with the
sign of the right-hand side of Eq.~(\ref{cascade2_3S1S}).

In the sum of the radiative corrections to one-photon decay and 
the two-photon decay, the incoherent cascade contributions cancel,
and this is in analogy to the discussion in Ref.~\cite{Ri1972}
for a different, but physically related process, namely the 
coherent/incoherent pair production via a virtual/real photon
intermediate state by an electron in crossed, static electromagnetic fields.

Immediately, new questions arise.
Our considerations suggest that our formulation in Eq.~(\ref{f}) 
provides infinitesimal imaginary parts that are appropriate for the 
evaluation of radiative corrections to the one-photon decay,
but provides the ``wrong'' sign of the cascades for two-photon decays.
This could lead to new doubt regarding whether we can extract a valid expression
for the two-photon decay rate from our Eq.~(\ref{f}) in the 
first place. The question is: Can we extract, by some mathematically 
justifiable procedure, from Eq.~(\ref{decay_length_3S1S_modulus}),
an expression for the two-photon decay rate which either confirms 
or invalidates our result for the two-photon decay rate, under a 
suitable gauge-invariant subtraction of the cascade contribution from the 
integrand in (\ref{decay_length_3S1S_modulus})?

First, since the cascade contributions correspond to the 
delta function in Eq.~(\ref{formula}), it is clear that the two-photon 
decay rate corresponds to the product of two principal-value 
distributions of the form, 
\begin{equation}
\left( {\rm P.V.} \frac{1}{\omega - \omega_0} \right)\,
\left( {\rm P.V.} \frac{1}{\omega - \omega_0} \right) = 
\left( {\rm P.V.} \frac{1}{\omega - \omega_0} \right)^2 \,,
\end{equation}
which is integrated over $\omega$.
As similar problems have occurred in field theory
[see Eq.~(6.23) on p.~168 of Ref.~\cite{BaKoSt1998}],
we are provided with a guiding principle
for the calculation. Namely, 
we consider an arbitrary function $f$, integrated
over a finite interval $(0,\omega_{\rm max})$ with 
$f(0) = f(\omega_{\rm max}) = 0$:
\begin{align}
\label{guidance}
\int_0^{\omega_{\rm max}} {\rm d}\omega \,
f(\omega) \, \left( {\rm P.V.} \frac{1}{\omega - \omega_0} \right)^2 = & 
\lim_{\eta \to 0} 
\int_0^{\omega_{\rm max}} {\rm d}\omega \,
f(\omega) \, \left( {\rm P.V.} \frac{1}{\omega - \omega_0 + \eta } \right) \,
\left( {\rm P.V.} \frac{1}{\omega - \omega_0 } \right) 
\nonumber\\[2ex]
= & 
\lim_{\eta \to 0} 
\frac{1}{\eta}
\int_0^{\omega_{\rm max}} {\rm d}\omega \,
f(\omega) \, \left( {\rm P.V.} \frac{1}{\omega - \omega_0 } -
{\rm P.V.} \frac{1}{\omega - \omega_0 + \eta} \right) 
\nonumber\\[2ex]
= & 
\lim_{\eta \to 0} 
\frac{1}{\eta}
\int_0^{\omega_{\rm max}} {\rm d}\omega \,
\left[ f(\omega) - f(\omega - \eta) \right] \, 
\left( {\rm P.V.} \frac{1}{\omega - \omega_0 } \right)
\nonumber\\[2ex]
= & 
\int_0^{\omega_{\rm max}} {\rm d}\omega \,
\left[ \frac{\partial}{\partial\omega} \left( f(\omega) - f(\omega_0) \right) \right] \, 
\left( {\rm P.V.} \frac{1}{\omega - \omega_0 } \right)
\nonumber\\[2ex]
= & 
-\frac{\omega_{\rm max} \, f(\omega_0)}
{\omega_0 \, (\omega_{\rm max} - \omega_0)} +
{\rm P.V.} \int_0^{\omega_{\rm max}} {\rm d}\omega \,
\frac{ f(\omega) - f(\omega_0) }{(\omega - \omega_0)^2 } \,.
\end{align}

This subtraction, applied to Eq.~(\ref{decay_length_3S1S_modulus}),
gives rise to 
\begin{align}
\label{renorm1}
& \Gamma^{(2)}_{3S} = \frac{4 \alpha^2}{27 \pi m^2} \,
{\rm P.V.} \!\!\!\!\!\!\!\!
\int\limits_0^{E_{3S} - E_{1S}}
 \!\!\!\!\!\!\!\!
{\rm d}\omega \, 
\left(
\omega^3 \, (E_{3S} - E_{1S} - \omega)^3 \, 
\left| \sum_{\nu}
\left\{ 
\frac{\left< 1S \left| \left| \vec{x} \right| \right| \nu P  \right> 
\left< \nu P \left| \left| \vec{x} \right| \right| 3S \right> }
  {E_{3S} - E_{\nu P}  - \omega} +
\frac{\left< 1S \left| \left| \vec{x} \right| \right| \nu P  \right> 
\left< \nu P \left| \left| \vec{x} \right| \right| 3S \right> }
  {E_{1S} - E_{\nu P} + \omega} \right\} \right|^2 
\right.
\nonumber\\[2ex]
 & \; \left.
- (E_{3S} - E_{2P})^3 \, (E_{2P} - E_{1S})^3 \, 
\left( \frac{ \left< 1S \left| \left| \vec{x} \right| \right| 2 P  \right>^2 
\left< 2 P \left| \left| \vec{x} \right| \right| 3S \right>^2 }
{(E_{3S} - E_{2P}  - \omega)^2} +
\frac{ \left< 1S \left| \left| \vec{x} \right| \right| 2 P  \right>^2 
\left< 2 P \left| \left| \vec{x} \right| \right| 3S \right>^2 }
{(E_{2P} - E_{1S}  - \omega)^2} \right) 
\right) + \frac{4 \alpha^2}{27 \pi m^2} \, {\cal F} \,.
\end{align}
Here, we have subtract the cascade-generating terms according to 
Eq.~(\ref{guidance}), thus leading to an integral which is 
finite under a principal-value prescription, because the 
double poles have explicitly been subtracted. Because the prescription
(\ref{guidance}) takes the numerators to exact resonance, the 
subtraction terms in (\ref{renorm1}) are gauge invariant and indeed
proportional to $\Gamma^{(1)}_{3S \to 2P} \, \Gamma^{(1)}_{2P \to 1S}$.
The additional term $\mathcal F$ is due to the boundary term 
found in (\ref{guidance}), 
\begin{align}
\label{renorm2}
{\mathcal F} =& 
- 2 (E_{3S} - E_{1S}) \,  (E_{3S} - E_{2P})^2 \, (E_{2P} - E_{1S})^2 \, 
\left< 1S \left| \left| \vec{x} \right| \right| 2 P  \right>^2 
\left< 2 P \left| \left| \vec{x} \right| \right| 3S \right>^2  \,.
\end{align}
Finally, returning to our original ${\rm i}\epsilon$
prescription, we have according to Eq.~(\ref{model}),
\begin{equation}
\label{model2}
\lim_{\epsilon \to 0} {\rm Re}
\int\limits_0^{\omega_{\rm max}} {\rm d}\omega \, 
\frac{f(\omega_0)}{(\omega - \omega_0 + {\rm i}\epsilon)^2} = 
-\frac{\omega_{\rm max} \, f(\omega_0)}
{\omega_0 \, (\omega_{\rm max} - \omega_0)} 
\end{equation}
and in view of (\ref{guidance}) and (\ref{model2}),
we obtain the (perhaps somewhat surprising) equality 
\begin{equation}
\lim_{\epsilon \to 0} {\rm Re}
\int\limits_0^{\omega_{\rm max}} {\rm d}\omega \, 
\frac{f(\omega)}{(\omega - \omega_0 + {\rm i}\epsilon)^2} = 
\int\limits_0^{\omega_{\rm max}} {\rm d}\omega \, 
f(\omega) \left( {\rm P.V.} \frac{1}{\omega - \omega_0} \right)^2 \,,
\end{equation}
which is subject to the interpretation
of the squared principal-value contribution according to  Eq.~(\ref{guidance}).
We can finally state that the result (\ref{renorm1}) agrees with 
formula (\ref{decay_length_3S1S}), so that, under the provisions
of the regularization implied by Eq.~(\ref{guidance}),
it is irrelevant if we start from an expression where the integrand 
for the two-photon decay is formulated as a modulus squared or with 
two infinitesimal imaginary parts ``pointing in the same direction.''

%
%
\section{Conclusions}
\label{conclusions}

We have analyzed two-photon decay processes
involving $nS \to 1S$ and $nD \to 1S$ channels in
hydrogenlike ions. Our general formulas
(\ref{decay_velocity}) and
(\ref{decay_length}) are gauge-invariant
and are obtained
with otherwise unspecified, arbitrary infinitesimal
imaginary parts ${\rm i}\, \epsilon$, provided the
limit $\epsilon \to 0$ is taken after the integrations
over the photon energies have been performed
(non-uniform convergence).
Numerical results are presented in Table~\ref{results}.
These are nonrelativistic results which scale
as $Z^6$ with the nuclear charge number $Z$.
For a relativistic generalization, see~\cite{SuJe2008}.

From a more philosophical point of view, we can say
that the two-photon decay process turns out to be 
an extremely subtle physical 
phenomenon, which demands a lot of mathematical sophistication 
in its analysis. Without a careful handling of the 
distributions, including ill-defined squares of delta 
functions, it is impossible to obtain consistent 
answers. The current work attempts to provide a proposal
for a consistent framework in which the resonant
intermediate states and the generated double poles 
can be addressed, while preserving the 
interpretation of the integrand of the two-photon 
decay rate as a differential decay rate with respect
to the photon energy.

Three remarks conclude this work. (i) Following Ref.~\cite{Ri1972}, 
we should point out that there is no guarantee that the 
coherent two-photon decay rate as evaluated here always needs 
to be positive (except for the $2S$ state,
where no resonant intermediate states are present). 
Indeed, as Eq.~(\ref{renorm1}) shows, the result is obtained 
as a subtracted integral, and the integrand is not necessarily 
positive. For all transitions considered here, the rate is positive
(see Table~\ref{results}),
but it is known that radiative corrections to decay rates can be negative, 
and the coherent two-particle contribution to a decay rate beyond the 
cascade constitutes a correction to a decay rate which need not 
be positive. This statement is paradoxical, but we can point out that 
this statement has already been confirmed after Eq.~(20) of Ref.~\cite{Ri1972}
in an absolutely analogous situation.
(ii) The observation time $T$ as implied in Eq.~(\ref{cascade2_3S1S}) 
has to be sufficiently large (larger than the typical formation 
time of radiation in the system, according to Ref.~\cite{Ri1972}, 
or otherwise the decay process will proceed in a different way).
In our case, the natural formation time of radiation is given by a time 
inversely related to the decay width of the decaying states,
which is naturally identified as the one-photon decay rate of the 
highly excited states.
(iii) It may seem that the agreement of the integration around the 
infinitesimally displaced poles as described in Sec.~\ref{gauge_invariance}
and the regularized principal-value prescription as 
described in Sec.~\ref{cascades} is purely accidental. However, one should
remember that similar integrals appear in Lamb-shift related 
self-energy calculations (e.g., Ref.~\cite{JeSoMo1997}),
and therefore, the predictions for the Lamb shift of excited states
would have had to be reinvestigated if we had not found 
agreement of the two computational schemes discussed here.
Fortunately, the internal consistency of mathematics protects us 
from having to reinvestigate accurate theoretical predictions
based on quantum electrodynamics.

\section*{Acknowledgment}

Helpful discussion with A. Surzhykov, A. I. Milstein, P. J. Mohr, 
F. Khan and K. Pachucki are gratefully acknowledged.
This work was supported by the Deutsche Forschungsgemeinschaft 
(Heisenberg program, contract JE285/3--1).


\begin{thebibliography}{10}

\bibitem{Je2007}
U.~D. Jentschura, J. Phys. A {\bf 40},  F223  (2007).

\bibitem{ChSu2007}
J. Chluba and R.~A. Sunyaev, e-print 0705.3033 [astro-ph].

\bibitem{Ri1972}
V.~I. Ritus, Nucl. Phys. B {\bf 44},  236  (1972).

\bibitem{BaKoSt1998}
V.~N. Baier, V.~M. Katkov, and V.~M. Strakhovenko, {\em Electromagnetic
  Processes At High Energies in Oriented Single Crystals} (World Scientific,
  Singapore, 1998).

\bibitem{Fl1984}
V. Florescu, Phys. Rev. A {\bf 30},  2441  (1984).

\bibitem{FlScMi1988}
V. Florescu, I. Schneider, and I.~N. Mihailescu, Phys. Rev. A {\bf 38},  2189
  (1988).

\bibitem{CrTaSaCh1986}
J.~D. Cresser, A.~Z. Tang, G.~J. Salamo, and F.~T. Chan, Phys. Rev. A {\bf 33},
   1677  (1986).

\bibitem{Pa2001}
K. Pachucki, Phys. Rev. A {\bf 63},  042503  (2001).

\bibitem{PaJe2003}
K. Pachucki and U.~D. Jentschura, Phys. Rev. Lett. {\bf 91},  113005  (2003).

\bibitem{BaSu1978}
R. Barbieri and J. Sucher, Nucl. Phys. B {\bf 134},  155  (1978).

\bibitem{LaLi1958}
L.~D. Landau and E.~M. Lifshitz, {\em Quantum Mechanics (Volume 3 of the Course
  of Theoretical Physics)} (Pergamon Press, London, 1958).

\bibitem{VaMoKh1988}
D.~A. Varshalovich, A.~N. Moskalev, and V.~K. Khersonskii, {\em Quantum Theory
  of Angular Momentum} (World Scientific, Singapore, 1988).

\bibitem{GM1931}
M. G\"{o}ppert-Mayer, Ann. Phys. (Leipzig) {\bf 9},  273  (1931).

\bibitem{BeSa1957}
H.~A. Bethe and E.~E. Salpeter, {\em Quantum Mechanics of One- and Two-Electron
  Atoms} (Springer, Berlin, 1957).

\bibitem{GaCo1970}
M. Gavrila and A. Costescu, Phys. Rev. A {\bf 2},  1752  (1970).

\bibitem{Pa1993}
K. Pachucki, Ann. Phys. (N.Y.) {\bf 226},  1  (1993).

\bibitem{JeEtAl2003}
U.~D. Jentschura, E.-O. Le~Bigot, P.~J. Mohr, P. Indelicato, and G. Soff, Phys.
  Rev. Lett. {\bf 90},  163001  (2003).

\bibitem{TuYeSaCh1984}
J.~H. Tung, X.~M. Ye, G.~J. Salamo, and F.~T. Chan, Phys. Rev. A {\bf 30},
  1175  (1984).

\bibitem{FlPaSt1987}
V. Florescu, S. Patrascu, and O. Stoican, Phys. Rev. A {\bf 36},  2155  (1987).

\bibitem{SaPaCh2004}
J. Sapirstein, K. Pachucki, and K.~T. Cheng, Phys. Rev. A {\bf 69},  022113
  (2004).

\bibitem{SuJe2008}
A. Surzhykov and U.~D. Jentschura, submitted (2008).

\bibitem{JeSoMo1997}
U.~D. Jentschura, G. Soff, and P.~J. Mohr, Phys. Rev. A {\bf 56},  1739
  (1997).

\end{thebibliography}
\end{document}